\documentclass[12pt,preprint]{aastex}
\newlength{\GBCdigit}
\newcommand{\GBC}{\hspace*{\GBCdigit}}
\newcommand{\grba}{GRB 020124}
\newcommand{\grbb}{GRB 030323}
\newcommand{\eiso}{E$_{\rm{iso}}$}
\newcommand{\epic}{E$_{\rm p}$}

\begin{document}

\shorttitle{HETE-2 Observation of two gamma-ray bursts}
\shortauthors{Atteia et al.}

\title{HETE-2 Observation of two gamma-ray bursts at z $> 3$}

\author{
J.-L.~Atteia,\altaffilmark{3}
N.~Kawai,\altaffilmark{4,5}
R.~Vanderspek,\altaffilmark{1}
G.~Pizzichini,\altaffilmark{16}
G.~R.~Ricker,\altaffilmark{1}
C.~Barraud,\altaffilmark{3}
M.~Boer,\altaffilmark{18}
J.~Braga,\altaffilmark{14}
N.~Butler,\altaffilmark{1}
T.~Cline,\altaffilmark{13}
G.~B.~Crew,\altaffilmark{1}
J.-P.~Dezalay,\altaffilmark{11}
T.~Q.~Donaghy,\altaffilmark{2}
J.~Doty,\altaffilmark{1}
E.~E.~Fenimore,\altaffilmark{10}
M.~Galassi,\altaffilmark{10}
C.~Graziani,\altaffilmark{2}
K.~Hurley,\altaffilmark{6}
J.~G.~Jernigan,\altaffilmark{6}
D.~Q.~Lamb,\altaffilmark{2}
A.~Levine,\altaffilmark{1}
R.~Manchanda,\altaffilmark{15}
F.~Martel,\altaffilmark{1}
M.~Matsuoka,\altaffilmark{8}
E.~Morgan,\altaffilmark{1}
Y. Nakagawa,\altaffilmark{9}
J.-F.~Olive,\altaffilmark{11}
G.~Prigozhin,\altaffilmark{1}
T.~Sakamoto,\altaffilmark{4,5,13}
R. Sato,\altaffilmark{4}
Y.~Shirasaki,\altaffilmark{5,7}
M.~Suzuki,\altaffilmark{4}
K.~Takagishi,\altaffilmark{17}
T.~Tamagawa,\altaffilmark{5}
K.~Torii,\altaffilmark{19}
J.~Villasenor,\altaffilmark{1}
S.~E.~Woosley,\altaffilmark{12}
M.~Yamauchi,\altaffilmark{17}
~and~A.~Yoshida\altaffilmark{5,9}
}

\altaffiltext{1}{Center for Space Research, Massachusetts Institute of
Technology, 70 Vassar Street, Cambridge, MA, 02139.}

\altaffiltext{2}{Department of Astronomy and Astrophysics, University
of Chicago, 5640 South Ellis Avenue, Chicago, IL 60637.}

\altaffiltext{3}{Laboratoire d'Astrophysique, Observatoire
Midi-Pyr\'{e}n\'{e}es, 14 Ave. E. Belin, 31400 Toulouse, France.}

\altaffiltext{4}{Department of Physics, Tokyo Institute of Technology, 
2-12-1 Ookayama, Meguro-ku, Tokyo 152-8551, Japan.}

\altaffiltext{5}{RIKEN (Institute of Physical and Chemical Research),
2-1 Hirosawa, Wako, Saitama 351-0198, Japan.}

\altaffiltext{6}{University of California at Berkeley,
Space Sciences Laboratory, Berkeley, CA, 94720-7450.}

\altaffiltext{7}{National Astronomical Observatory, Osawa 2-21-1,
Mitaka,  Tokyo 181-8588 Japan.}

\altaffiltext{8}{Tsukuba Space Center, National Space Development
Agency of Japan, Tsukuba, Ibaraki, 305-8505, Japan.}

\altaffiltext{9}{Department of Physics, Aoyama Gakuin University,
Chitosedai 6-16-1 Setagaya-ku, Tokyo 157-8572, Japan.}

\altaffiltext{10}{Los Alamos National Laboratory, P.O. Box 1663, Los 
Alamos, NM, 87545.}

\altaffiltext{11}{Centre d'Etude Spatiale des Rayonnements,
Observatoire Midi-Pyr\'{e}n\'{e}es,
9 Ave. du Colonel Roche, 31028 Toulouse Cedex 4, France.}

\altaffiltext{12}{Department of Astronomy and Astrophysics, University 
of California at Santa Cruz, 477 Clark Kerr Hall, Santa Cruz, CA
95064.}

\altaffiltext{13}{NASA Goddard Space Flight Center, Greenbelt, MD,
20771.}

\altaffiltext{14}{Instituto Nacional de Pesquisas Espaciais, Avenida
Dos Astronautas 1758, S\~ao Jos\'e dos Campos 12227-010, Brazil.}

\altaffiltext{15}{Department of Astronomy and Astrophysics, Tata 
Institute of Fundamental Research, Homi Bhabha Road, Mumbai, 400 005, 
India.}

\altaffiltext{16}{INAF/IASF Sezione di Bologna, via Piero Gobetti 101, 
40129 Bologna, Italy.}

\altaffiltext{17}{Faculty of engineering, Miyazaki University, Gakuen
Kibanadai Nishi, Miyazaki 889-2192, Japan.}

\altaffiltext{18}{Observatoire de Haute Provence,
04870 St. Michel l'Observatoire, France.}

\altaffiltext{19}{Department of Earth and Space Science,
Graduate School of Science,
Osaka University,
1-1 Machikaneyama, Toyonaka, Osaka 560-0043, Japan}

\begin{abstract}

\grba\ and \grbb\ constitute half the sample of gamma-ray bursts
with a measured redshift greater than 3.
This paper presents the temporal and spectral properties 
of these two gamma-ray bursts detected and localized
with HETE-2. 
While they have nearly identical redshifts (z=3.20 for GRB 020124, 
and z=3.37 for GRB 030323), these two GRBs span about an order of magnitude 
in fluence, thus sampling distinct regions of the GRB luminosity function.
The properties of these two bursts are compared with those of 
the bulk of the GRB population detected by HETE-2. We also discuss
the energetics of \grba\ and \grbb\ and show that they are 
compatible with the \epic - \eiso\ relation discovered by Amati et al. (2002).
Finally, we compute the maximum redshifts at which these bursts
could have been detected by HETE-2 and we address various issues 
connected with the detection and localization of high-z GRBs.

\end{abstract}


\section{Introduction}

High redshift gamma-ray bursts (GRBs) may become
useful beacons for the study of the young universe. 
Gamma-ray bursts are expected to be visible out to very large redshifts
(z=10-20, Lamb \& Reichart 2000), if indeed they are generated there, 
and offer the possibility to
probe the interstellar medium along the line of sight and to address 
important cosmological issues like the evolution of the star formation rate.

Before one can undertake such studies, however, it is important
to understand the intrinsic properties of gamma-ray bursts at high redshift.
There are only four GRBs with a measured redshift larger than z=3: 
GRB 971214 at z=3.42 \cite{kulk98}, which was localized by BeppoSAX; 
GRB 000131 at z=4.5 \cite{ande00}, which was localized with the IPN; and
GRB 020124 at z=3.20 \cite{hjor03} and GRB 030323 at z=3.37 \cite{vree04},
which were localized by HETE-2. 

This paper describes the HETE-2 observations of GRB 020124 and GRB 030323.
After a quick summary of the localization history of these bursts 
(section \ref{detection}), we present their spectral and temporal properties 
in section \ref{prompt} and compare them with those of the bulk 
of the GRB population.
Their energetics are discussed in section \ref{energetics}, where it is noted 
that GRB 030323 is the first GRB detected at large redshift 
which does not belong to the bright 
end of the GRB luminosity distribution.
In this section we also show that HETE-2 could have detected and localized GRB 020124
at a redshift of z=6.4 at least, and we discuss some necessary conditions
for the detection of soft, faint GRBs at high redshift.
Section \ref{afterhost} looks into the observations of the afterglows 
and host galaxies of these bursts.

\section{Detection, localization, and optical afterglow identification} 
\label{detection}

\subsection{GRB 020124}
\begin{figure}
\plottwo{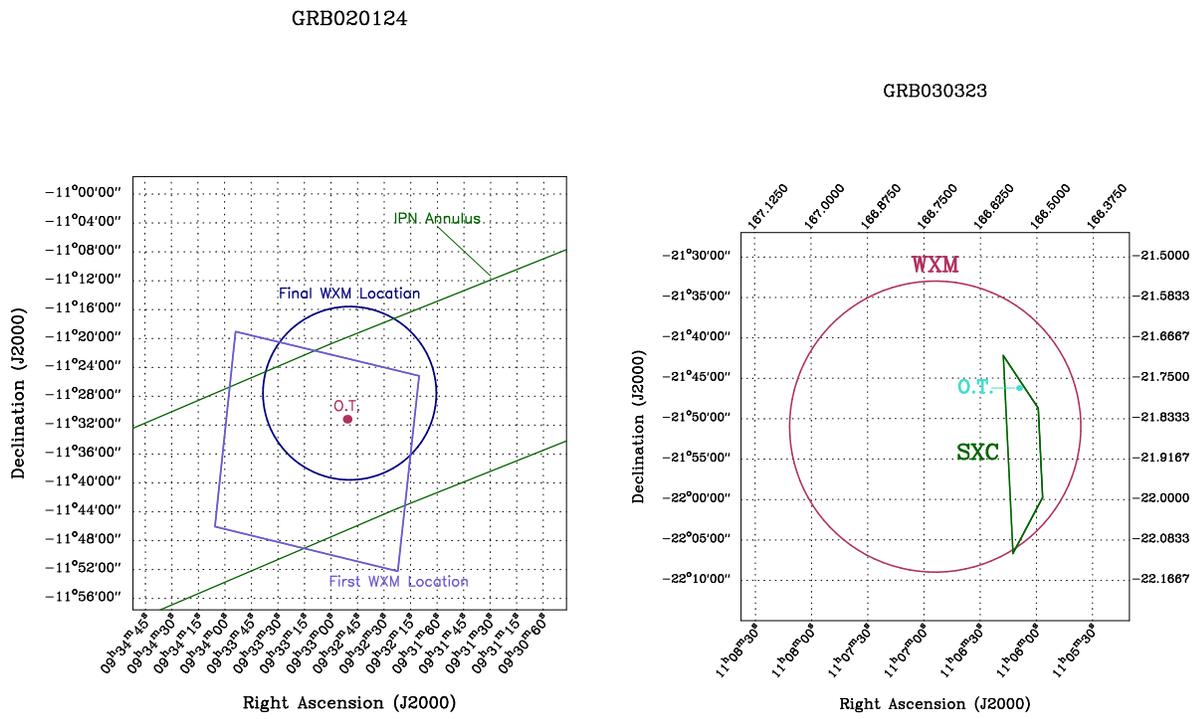}{f1b.eps}
\caption{Left panel: Reported localizations and optical afterglow of GRB 020124.
Right panel: Reported localizations and optical afterglow of GRB 030323.
\label{figloc}}
\end{figure}

On 2002 January 24, at 10:41:15.15 UT (38475.15 SOD), 
the High Energy Transient Explorer satellite HETE-2 
(hereafter HETE) detected GRB 020124, a moderately bright 
GRB. 
No flight localization was issued by the satellite, but the analysis
of data on the ground resulted in a coarse localization distributed to
the GCN\footnote{Gamma-Ray Burst Coordinate Network, see http://gcn.gsfc.nasa.gov/}
1.4 hours after the GRB, and in a refined position distributed
to the GCN 10.7 hours after the GRB (Ricker et al. 2002).
The refined position was a circle of radius 12', centered at 
RA = 09h 32m 49s,  Dec = -11$^\circ$ 27' 35'' (J2000).
The optical afterglow was reported 28 hours after the GRB, 
(in data taken 13.5 hours after the GRB) at the position:
RA = 09h 32m 50.8s,  Dec = -11$^\circ$ 31' 11'',  
well within the refined error box \cite{pric02}. 
An IPN annulus was also obtained for this burst, which was 
fully consistent with the WXM error box \cite{hurl02}.
Fig. \ref{figloc}a shows the projection of the WXM error box on the sky, 
the IPN annulus, and the position of the optical afterglow. 
At the time of its identification, the afterglow had a magnitude R=18.5
\cite{pric02}.
Details of the identification of the optical afterglow and its evolution
can be found in Berger et al. (2002). 
The afterglow spectrum recorded with ISAAC on the VLT-Antu is analysed
in Hjorth et al. (2003), who report a redshift z=3.198.

%

\subsection{GRB 030323}
On 2003 March 23, at 21:56:57.60 UT (79017.60 SOD), HETE detected GRB 030323, a faint GRB. 
As for \grba\ no flight localization was issued by the satellite, but the analysis
of data on the ground resulted in a WXM localization distributed to
the GCN 5.0 hours after the GRB, and in an SXC localization distributed 
to the GCN 7.5 hours after the GRB (Graziani et al. 2003).
The WXM position was a circle of radius 18', centered at 
RA = 11h 06m 54s,  Dec = -21$^\circ$ 51' 00'' (J2000).
The SXC position was a trapezoid with an area of 71 arcmin, 
fully included within the WXM error box. 
The center of the SXC error box was RA = 11h 06m 06s,  Dec = -21$^\circ$ 54' 20'' (J2000).
The optical afterglow was reported 22 hours after the GRB, 
(in data taken 9.6 hours after the GRB), at the position:
RA = 11h 06m 09.38s,  Dec = -21$^\circ$ 46' 13.3'' , 
close to the boundary of the SXC error box, but within it \cite{gilm03}. 
Fig. \ref{figloc}b shows the projection of the WXM and SXC
error boxes on the sky, and the position of the optical afterglow. 
At the time of its identification, the afterglow had a
magnitude Rc=18.7 \cite{gilm03}.
A temporal and spectral analysis of the optical afterglow can be found in
Vreeswijk et al. (2004), who report a redshift z=3.372.

%

\section{The prompt emission} \label{prompt}

\subsection{Light curves and temporal properties} \label{lightcurve}

\begin{figure}
\plottwo{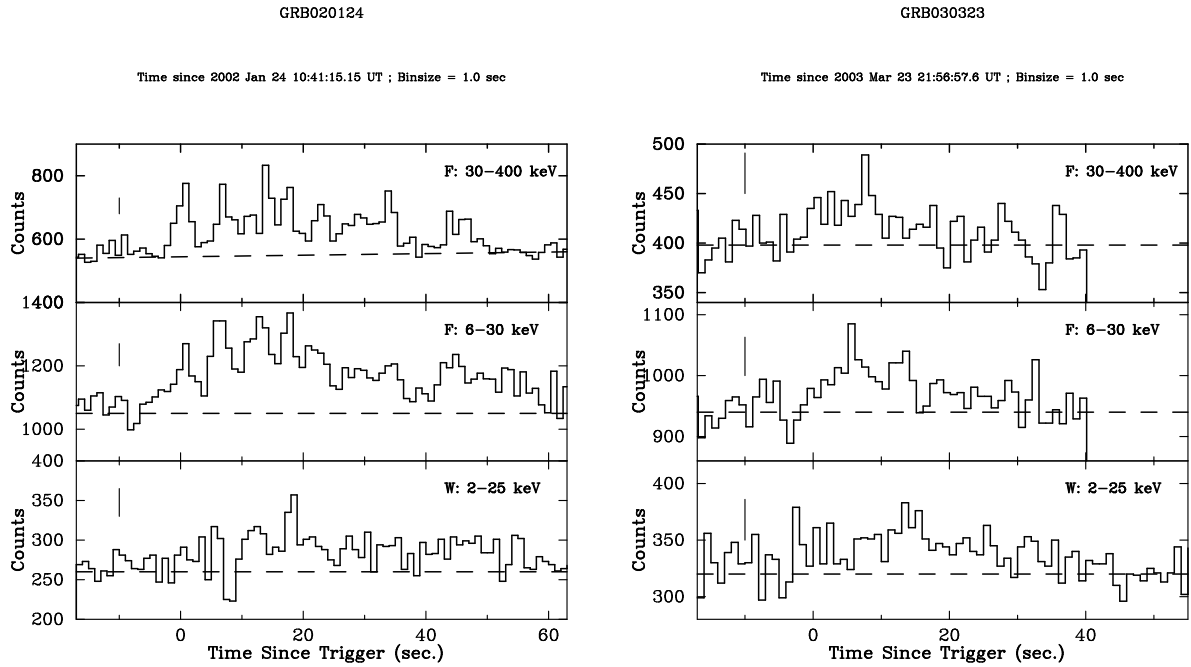}{f2b.eps}
\caption{Left panel: Light curves of GRB 020124 in various energy bands.
The lower panel shows the light curve recorded with the WXM (2-25 keV).
The upper two panels show the light curves recorded with FREGATE 
in two energy bands (6-30 keV, and 30-400 keV). The vertical line 
at t=-10 sec shows the typical size of 1-sigma error bars.
Right panel: Same plots for GRB 030323.
\label{figlc}}
\end{figure}

The light curves of GRB 020124 and GRB 030323 are displayed
in Fig. \ref{figlc}.
GRB 020124 and GRB 030323 belong to the class of long GRBs 
with durations T$_{90}$ = 49.4 $\pm 1.3$ sec (GRB 020124), and  
T$_{90}$ = 15.0 $\pm 2.6$ sec (GRB 030323) in the energy range 6-400 keV.
Table \ref{tabdur} gives the durations of these two bursts, 
and their one sigma errors, in various energy bands. 
In the FREGATE data, GRB 020124 exhibits little duration shortening 
with the energy with T$_{90}$ (resp. T$_{50}$) varying from 51.4 $\pm$ 1.4 sec (resp.
26.0 $\pm$ 2.7 sec) in the energy band 6-15 keV to 43.0 $\pm$ 6.1 sec
(resp. 19.2 $\pm$ 1.8 sec) in the energy range 85-400 keV.
The situation is more confusing in the WXM 2$-$25 keV energy band
(T$_{90}$ and T$_{50}$ seem
to follow distinct trends), possibly due to the lower signal to noise ratio
of the burst in this instrument.
The light curve of GRB 020124 appears very spiky: at least 9
individual spikes can be identified in it.

GRB 030323 is significantly fainter than GRB 020124, and we could only
divide the energy range into two subranges : in 6-30 keV we measure
 T$_{90}$ = 12.8 $\pm$ 2.5 sec and  T$_{50}$ = 6.6 $\pm$ 1.5 sec, and 
in 30-400 keV we measure  T$_{90}$ = 12.2 $\pm$ 3.6 sec and  T$_{50}$ = 5.2 $\pm$ 1.6 sec.
With only two energy bands, it is difficult to say whether GRB 030323
exhibits significant duration shortening with energy.

\subsection{Spectra} \label{spectrum}

In this section we investigate the average spectral properties of \grba\ and \grbb .
The joint WXM+FREGATE spectra have been fit with a powerlaw times exponential model
(PLE), $n(E) \propto E^{-\alpha} \exp ( -E / e_0^{\rm obs} )$, 
and with a Band function (GRBM), which
satisfactorily fits most GRB spectra \cite{band93}.\footnote{In this paper we use 
capital letters for intrinsic spectral parameters (at the source), and lower case
letters for the observed spectral parameters.}
This parametrization allows us to compare these bursts with the GRBs detected
by BATSE \cite{band93,pree00}, BeppoSAX \cite{fron00,amat02}, and 
HETE \cite{barr03,saka04b}.
The spectral parameters for the burst-averaged spectra of GRB 020124 and GRB 030323
are given in Table \ref{tabspectrum}.
The emission properties of GRB 020124 and GRB 030323 are given in Table \ref{tabemission}.
Emission properties depend on the model used in the spectral analysis.
In Table \ref{tabemission} we report the numbers given in Sakamoto et al. (2004b), 
which are based on a PLE fit for GRB 020124 and on a simple powerlaw fit for GRB 030323. 
It should be noted that a spectral analysis based on the best fit Band function (given 
in Table \ref{tabspectrum}) gives emission parameters 
which differ by less than 10\% from the values given in Table \ref{tabemission}. 
The discussion in section \ref{energetics}, which requires the measure of
'bolometric' fluences is based on the best fit Band functions given in 
Table  \ref{tabspectrum}.

\subsubsection{GRB 020124}
With a fluence of $8.1 \times 10^{-6}$ erg cm$^{-2}$ in the energy range 
2 - 400 keV, \grba\ belongs to the 
brightest third of HETE GRBs. 
Its 'softness'\footnote{Following \cite{saka04a}, we define
the softness as the ratio $S_x/S_{\gamma}$,where $S_x$ is 
the fluence in the range 2-30 keV, 
and $S_\gamma$ is the fluence in the range 30-400 keV. 
Using this definition X-Ray Flashes have a softness greater 
than 1 and X-Ray Rich GRBs have a softness in the range 0.33 to 1.}
is 0.32, placing it at the boundary between GRBs and X-Ray Rich GRBs.
The peak energy, e$_{\rm p}^{\rm obs}$ , is relatively well constrained 
to be about 90 keV.
The difference between this value and the value of 133 keV quoted in Barraud et al. (2003)
is due to the inclusion of the WXM data covering the range 2-25 keV 
in this refined analysis. 

\subsubsection{GRB 030323}
With a fluence of $1.2 \times 10^{-6}$ erg cm$^{-2}$ in the energy range 
2 - 400 keV, \grbb\ belongs to the 
faintest 20\% of HETE GRBs. 
Its 'softness' is 0.38, making it an X-Ray Rich GRB.
The e$_{\rm p}^{\rm obs}$ value of $\sim 60$ keV is not well constrained and could be
as low as 20 keV or as high as 200 keV.

\section{Distance and energetics} \label{energetics}

Knowing the redshifts of \grba\ and \grbb\ allows us to compute their
{\it intrinsic} spectral parameters: E$_{\mathrm p}$, the peak energy
of the $\nu$F$\nu$ spectrum; \eiso , the isotropic-equivalent
radiated energy (in the energy range 1-10000 keV), and 
N$_\gamma$ the isotropic-equivalent photon number in the same energy range.
In the following we adopt a flat cosmology with $\Omega_m = 0.3$,
$\Omega_{\Lambda} = 0.7$, and h$_0 = 0.65$.
For GRB 020124, we get \epic\ = $390^{+70}_{-120}$ keV, 
E$_{\mathrm{iso}} = 25\pm 6 \times 10^{52}$ erg, 
and N$_\gamma = 34^{+20}_{-10} \times 10^{58}$ photons.
For GRB 030323, we get \epic\ = $270^{+600}_{-180}$ keV, 
E$_{\mathrm{iso}} = 3.2\pm 1 \times 10^{52}$ erg,
and N$_\gamma = 5.4^{+6}_{-2.5} \times 10^{58}$ photons.


We note that GRB 020124 is intrinsically bright, 
similar to GRB 971214, which had \eiso = $3 \times 10^{53}$ erg 
\cite{kulk98,dalf00,amat02}; 
and GRB 000131 which had  \eiso = $11 \times 10^{53}$ erg \cite{ande00}.
The isotropic-equivalent energy of GRB 030323
is however about an order of magnitude fainter, a fact that demonstrates
the ability of HETE to detect and localize high-redshift GRBs that
are not at the bright end of the GRB luminosity function.

We have checked whether \grba\ and \grbb\ are compatible with the empirical 
\eiso - \epic\ relation discovered by Amati et al. (2002), which can be expressed 
as E$_{\rm{iso}}^{0.5}$ / \epic $\sim$ 1 (where \eiso\ is measured in units 
of 10$^{50}$ erg and \epic\ in keV; see also Lamb et al. 2005)
This is indeed the case with E$_{\rm{iso}}^{0.5}$ / \epic  = 1.3 for GRB 020124. 
For GRB 030323 the \epic\ of \grbb\ is not well determined and we cannot draw any
conclusion based on this burst (although for the sake of completeness 
we should mention that the best fit values give E$_{\rm{iso}}^{0.5}$ / \epic = 0.65).
The spectral parameters of \grba\ are well determined, and the fact
that this burst follows the \eiso - \epic\ relation closely could
indicate that this relation has little or no evolution with the redshift.



\subsection{The maximum distances at which \grba\ and \grbb\ could be localized} \label{dmax}

Fig. \ref{softness} shows the position of \grba\ and \grbb\ 
(the two large triangles, \grba\ is the rightmost large triangle) in a fluence-softness
diagram, among the population of GRBs detected with HETE (Barraud et al. 2004). 
This figure also shows the positions that these bursts would have in the same
diagram if they had occured at redshifts 1, 2, 5, 10, and 20. 
The tracks of \grba\ and \grbb\ in the figure are computed by 
taking into account the spectral redshift of the bursts, and
assuming that we can measure the total fluence, even when the 
burst is at a high redshift. 
From this figure we can see that, at redshift 1, \grba\ would have been 
one of the brightest GRBs detected by HETE, while \grbb\ would have
been in the middle of the fluence distribution.
Regarding their hardness, we note that at a redshift of unity, \grba\
and \grbb\ would both have been unambiguously classified as 'GRBs', 
rather than XRR-GRBs or XRFs. 

Figure \ref{softness} can also be used to discuss 
the maximum redshift at which
HETE could have detected \grba\ and \grbb .
Based on this figure, we see that \grbb\ is very close 
to the boundary of the GRB population detected and localized with HETE.
In contrast, the fluence of \grba\ appears to remain well 
within the distribution of localized GRBs up to redshift z $\sim$ 15.
In reality, the fluence is not the best intensity indicator 
for GRB {\it detection} because the trigger algorithm is essentially
based on the search for excess counts in short time intervals.
In order to assess more precisely the maximum redshift at which
HETE could have detected \grba , we have performed 
a detailed analysis including the following steps:

- Compute the trigger threshold (number of counts) of FREGATE 
at the time of \grba , on the 1.3 sec and on the 5.2 sec trigger timescales.

- Estimate the signal to noise ratio of \grba\ for its detection by FREGATE.

- Determine the redshift at which the counts from \grba\ would reach the 
trigger threshold of FREGATE, taking into account the effects of the distance, 
of the time dilation and of the spectral redshift of the photons.

This analysis shows that a burst 3.2 times fainter than \grba\ would 
have triggered FREGATE in the 7-30 keV energy range in a time window 
of 5.2 sec. We thus conclude that \grba\ could have been 
detected by FREGATE up to a redshift z=6.4.
The trigger scheme of HETE includes many more trigger possibilities 
than the simple trigger palette of FREGATE, especially some using 
longer timescales for triggering. It is thus reasonable to assume that
\grba\ could have been detected by HETE up to redshift 7-8.
At this redshift HETE could have localized GRB 020124 
because its localization capabilities are more dependent on the fluence
of the burst than on its peak flux, and because the fluence decreases
more slowly than the peak flux with the redshift.
This is confirmed by figure \ref{softness} which shows that,
at a redshift z=8, the fluence and 
the softness of GRB 020124 would have been comparable with the fluences 
and softness of many GRBs localized by HETE. 

We finally note that, at redshifts higher than 10, 
the fluence of \grba\ remains comparable with
the fluence of many GRBs detected and localized with HETE
while its peak flux is well below the detection threshold.
A consequence of this fact is that a mission can greatly 
improve its ability
to detect high-z GRBs if it is able to localize long, faint,
soft transients. This is not the case for BeppoSAX and HETE-2 whose
detection strategy is mainly based on the search for count excesses on
relatively short timescales.
A strategy based on the search for
long, soft transients appearing in the {\it image} of the sky, 
as is the case for SWIFT-BAT, appears more promising for detecting high-z GRBs.

\section{Afterglows and hosts of \grba\ and \grbb\ } \label{afterhost}

The discussion in this section is mainly based on the information that has been
published in papers 
on the afterglows and hosts of the four GRBs with a measured redshift 
greater than 3: GRB 971214 \cite{halp98,kulk98,dalf00},
GRB 000131 \cite{ande00}, GRB 020124 \cite{berg02,hjor03}, and
GRB 030323 \cite{vree04}.

\subsubsection{Spectroscopy of the afterglows} \label{spectroscopy}

Thanks to fast localizations by HETE, the afterglows of GRBs 020124 and 030323
were observed soon after the burst, while they were still bright.
The identification of the afterglow of \grba\ is due to Price et al. (2002)
in data taken 13.5 hours after the burst, but the first observation 
took place only two hours after the trigger and caught the afterglow
at a magnitude R=18.5 \cite{tori02}.
The identification of the afterglow of GRB 030323 is due to 
Gilmore et al. (2003) in data taken 9.5 hours after the burst.
The afterglow was first observed with ROTSE III, when it had a
magnitude R=18.4, 6 hours after the burst. 
This contrasts with the afterglows of GRB 971214, and GRB 000131,
which were identified when they had magnitudes R=22.1,
and R=23.3, respectively.  
These quick detections allowed for spectroscopy of the afterglows
while they were still moderately bright (R=24 for \grba\ and R=21.5 for 
\grbb ). 
The spectroscopic observations of \grba\ and \grbb\ resulted in high
quality spectra and in the detection of strong Absorption Line Systems
(ALS) in the afterglows of both GRBs \cite{hjor03,vree04}.

\subsubsection{Beaming breaks} \label{breaks}
The light curves of GRB afterglows often display jet breaks, attributed to the 
confinement of the relativistic outflow into a small cone.
Rhoads (1997) has shown that jet breaks can be used 
to estimate $\theta_j$, the jet opening angle, and 
E$_\gamma$, the total energy output in $\gamma$-rays.
E$_\gamma$ is given by the following formula

$$\mathrm E_\gamma = E_{iso} * ({1 - cos(\theta_j)})$$

\noindent where $\theta_j$ is the opening angle of the jet in radians.

The afterglows of \grba\ and \grbb\ were observed on several occasions
during the few days following the bursts, but not frequently enough
to unambiguously identify a possible jet break.
Berger et al. (2002) nevertheless argue that \grba\ might 
have a jet break 10-20 days after the burst.
The observations of \grbb , on the other hand, give an afterglow slope  
s=$-1.56 \pm 0.03$ \cite{vree04}, showing that they 
took place before a possible jet break (one expects s $\sim -2$
after the break).
Fig.1. of Vreeswijk et al. (2004) shows that we can exclude 
a jet break in the afterglow, in the first 4 days following the burst.

Frail et al. (2001) find that E$_\gamma$ is narrowly distributed
around $5 \times 10^{50}$ erg in a sample of GRBs with known redshifts. 
Bloom, Frail, and Kulkarni (2003) later revised this value to
$1.3 \times 10^{51}$ erg.
From the measured value of \eiso , we can calculate the opening 
angle that the jet would have to have for E$_\gamma$ to be $1.3 \times 10^{51}$ erg.
For \grba , we find $\theta_j$ = 0.10 radians (5.8 degrees), and 
$\theta_j$ = 0.29 radians (16 degrees) for \grbb .
Following equation (1) of Frail et al. (2001), and assuming 
a circumburst density of [0.1 cm$^{-3}$], we find that the expected
break time is 2.6 days for \grba , and 5.7 days for \grbb .
With the freedom allowed by the poor sampling of the light curves,
and by the small (but real) scatter in the size of the energy reservoir, 
we consider that these numbers 
do not contradict the finding of Frail et al. (2001), and 
Bloom, Frail, and Kulkarni (2003) that there is a standard 
radiated energy for GRBs.

\subsubsection{Host galaxy identification} \label{hosts} 
\grba\ and \grbb\ have very faint hosts: R $\ge$ 29.5 for \grba\
\cite{bloo02c}, and V=28.0 for \grbb\ \cite{vree04}.
The early localization of these two GRBs, and the quick identification
of their afterglows, made possible the identification of the host 
galaxy of GRB 030323, and 
placed stringent limits on the magnitude of the host galaxy of GRB 020124.
The faintness of the host galaxies of these two bursts shows the
impossibility to measure their redshifts from the spectroscopy of their 
host galaxies.
\grba\ and \grbb\ are examples of GRBs whose redshifts can only
be measured at early times from the spectrum of the afterglow 
(unlike what was done for GRB 971214).
The fact that two of the four GRBs known with z $\ge$ 3 occurred
in faint galaxies, may indicate that a non-negligible fraction of star
formation takes place in such faint galaxies. Gamma-ray bursts 
appear to be a privileged way to identify this population.












\section{Conclusions} \label{conclusions}

This paper describes the temporal and spectral properties 
of \grba\ and \grbb , two GRBs at redshift z $>$ 3 detected and localized with HETE.
These two events are found to be fully consistent with the properties of the
rest of the GRB population detected with HETE.

We have used the chain of events which successfully led 
to the measurement of the redshifts of \grba\ and \grbb\ as a
baseline to discuss the conditions required for the identification of high-z GRBs.  
Our two main conclusions are summarized below.

The fast localization of \grba\ and \grbb\ allowed the quick identification
and the early spectroscopy of their afterglows. 
We see a posteriori that this was the only way to measure their redshifts,
given the faintness of their host galaxies.
In these cases, contrary to the case of GRB 971214, we could not rely
on the spectroscopy of the host galaxy to measure the redshifts, and
this might well be the case for the majority of high-z GRBs.
 
Study of  \grba\ shows that even instruments of modest size like FREGATE
or the WXM are able to detect and localize GRBs up to z=7-8, if indeed
GRBs occur at these redshifts.
The study of the tracks with redshift of the peak flux and of the fluence 
of \grba\ provides insight into the strategy to be used for
the detection of high-z GRBs. 
A strategy that relies mainly on the search for count excesses in
short time intervals does not appear to be the most appropriate. A strategy based 
on the imaging of faint, soft transients lasting minutes appears 
more promising.

\acknowledgments
\section*{Acknowledgments}

The HETE mission is supported in the U.S. by NASA contract NASW-4690; in
Japan, in part by the Ministry of Education, Culture, Sports, Science,
and Technology Grant-in-Aid 13440063; and in France, by CNES contract
793-01-8479.  KH is grateful for HETE support under Contract
MIT-SC-R-293291, for Ulysses support under JPL Contract 958056, and for
IPN support under NASA grant FDNAG5-11451.  G. Pizzichini acknowledges
support by the Italian Space Agency.



\clearpage



\begin{deluxetable}{lccccc}
\tablecaption{Temporal Properties of GRB 020124 and GRB 030323.
\label{tabdur}}
\tablewidth{0pt}
\tablehead{
\colhead{GRB / Instrument} & \colhead{Energy} & \colhead{$t_{90}$} & \colhead{$t_{50}$} \\
& \colhead{(keV)} & \colhead{(s)} & \colhead{(s)}
}
\startdata

GRB 020124 &&& \\
HETE WXM     & \GBC2--25\GBC    & 50.2 $\pm$ 2.3 & 18.6 $\pm$ 1.1 \\
             & \GBC2--5\GBC     & 41.8 $\pm$ 0.4 & 23.5 $\pm$ 1.7 \\
             & \GBC5--10\GBC    & 50.4 $\pm$ 8.0 & 11.7 $\pm$ 3.0 \\
             & 10--25\GBC       & 32.5 $\pm$ 1.2 & 16.7 $\pm$ 3.5 \\
HETE FREGATE & \GBC6--400\GBC   & 49.4 $\pm$ 1.3 & 22.6 $\pm$ 1.0 \\
             & 6--15\GBC        & 51.4 $\pm$ 1.4 & 26.0 $\pm$ 2.7 \\
             & 15--30\GBC       & 52.9 $\pm$ 2.0 & 23.2 $\pm$ 2.4 \\
             & 30--85\GBC       & 45.6 $\pm$ 0.7 & 22.2 $\pm$ 1.3 \\
             & 85--400          & 43.0 $\pm$ 6.1 & 19.2 $\pm$ 1.8 \\
&&& \\
GRB 030323 &&& \\
HETE WXM     & \GBC2--25\GBC    & 32.6 $\pm$ 2.7 & 13.9 $\pm$ 1.6 \\
             & \GBC2--5\GBC     & 31.5 $\pm$ 0.9 & 16.2 $\pm$ 0.8 \\
             & \GBC5--10\GBC    & 36.1 $\pm$ 0.6 & 19.4 $\pm$ 1.3 \\
             & 10--25\GBC       & 19.5 $\pm$ 2.0 & 12.5 $\pm$ 1.3 \\
HETE FREGATE & \GBC6--400\GBC   & 15.0 $\pm$ 2.6 & 7.1 $\pm$ 1.2 \\
             & 6--30\GBC        & 12.8 $\pm$ 2.5 & 6.6 $\pm$ 1.5 \\
             & 30--400\GBC      & 12.2 $\pm$ 3.6 & 5.2 $\pm$ 1.6 \\
\enddata
\vskip -18pt
\tablecomments{Errors are 1-$\sigma$. 
The significantly longer duration measured by the WXM for GRB 030323
is explained by the better sensitivity of this instrument to low
energy photons which are hardly detected with FREGATE.}
\end{deluxetable}


\begin{deluxetable}{lccc}
\tablecaption{Spectral models for GRB020124 and GRB 030323.
See section \ref{spectrum} for a description of the spectral models.
\label{tabspectrum}}

\tablewidth{0pt}
\tablehead{
\colhead{GRB / Model} & \colhead{Alpha} & \colhead{E$_0$} & \colhead{Beta}
}
\startdata
&&& \\
GRB 020124 PLE  
& $-0.79^{+0.15}_{-0.14}$ & 87$^{+34}_{-21}$ &  N/A \\
&&& \\
GRB 020124 GRBM 
& $-0.87^{-0.16}_{+0.19}$ & 82$^{+31}_{-31}$ & $-2.6^{}_{-0.65}$ \\
&&& \\
GRB 030323 PLE 
& $-0.80^{+0.8}_{-0.83}$ & 44$^{+90}_{-26}$ &  N/A \\
&&& \\
GRB 030323 GRBM$^a$ 
& $-0.96^{+1.31}_{-0.85}$ & 60$^{}_{-45}$ & $-2.3$ (frozen) \\
&&& \\
\enddata
\vskip -18pt
\tablecomments{Errors are for 90\% confidence, there is no upper limit for Beta.
$^a$ For this model, the parameters are poorly constrained. 
$\beta$ is frozen at $-2.3$, and there is no upper limit on {E$_0$}.}
\end{deluxetable}

\begin{deluxetable}{rcccc}
\tablecaption{Emission Properties of GRB 020124 and GRB 030323.
\label{tabemission}}
\tablewidth{0pt}
\tablehead{
\colhead{Energy} &
\colhead{Peak photon Flux} &
\colhead{Peak energy Flux} & 
\colhead{Energy Fluence} \\
\colhead{(keV)} & \colhead{(ph cm$^{-2}$ s$^{-1}$)} & 
\colhead{(erg cm$^{-2}$ s$^{-1}$)} & \colhead{(erg cm$^{-2}$)}
}
\startdata
GRB 020124 &&& \\
2--30       & $6.9 \pm 1.6$ & $ 1.8 \pm 0.18 \times 10^{-7}$ & 
                                         $2.0^{+0.14}_{-0.14} \times 10^{-6}$ \\
30--400     & $2.5 \pm 0.40$ & $ 4.5 \pm 0.46 \times 10^{-7}$ & 
                                         $6.1^{+0.88}_{-0.76} \times 10^{-6}$ \\
50--300     & $1.4 \pm 0.28$ & $ 3.2 \pm 0.33 \times 10^{-7}$ & 
                                         $4.7^{+0.82}_{-0.82} \times 10^{-6}$ \\

&&& \\
GRB 030323 &&& \\
2--30       & $3.4 \pm 2.1$ & $ .57 \pm .16 \times 10^{-7}$ & 
                                         $3.4^{+1.3}_{-1.2} \times 10^{-7}$ \\
30--400     & $0.49 \pm 0.22$ & $ 1.5 \pm .43 \times 10^{-7}$ & 
                                         $8.9^{+3.8}_{-3.5} \times 10^{-7}$ \\
50--300     & $0.29 \pm 0.15$ & $ 1.0 \pm .29 \times 10^{-7}$ & 
                                         $6.5^{+2.8}_{-2.8} \times 10^{-7}$ \\

\enddata
\vskip -18pt
\tablecomments{Errors are given at the 90\% confidence level.}
\end{deluxetable}

\clearpage

\clearpage

\begin{figure}
\includegraphics[angle=270,scale=0.25]{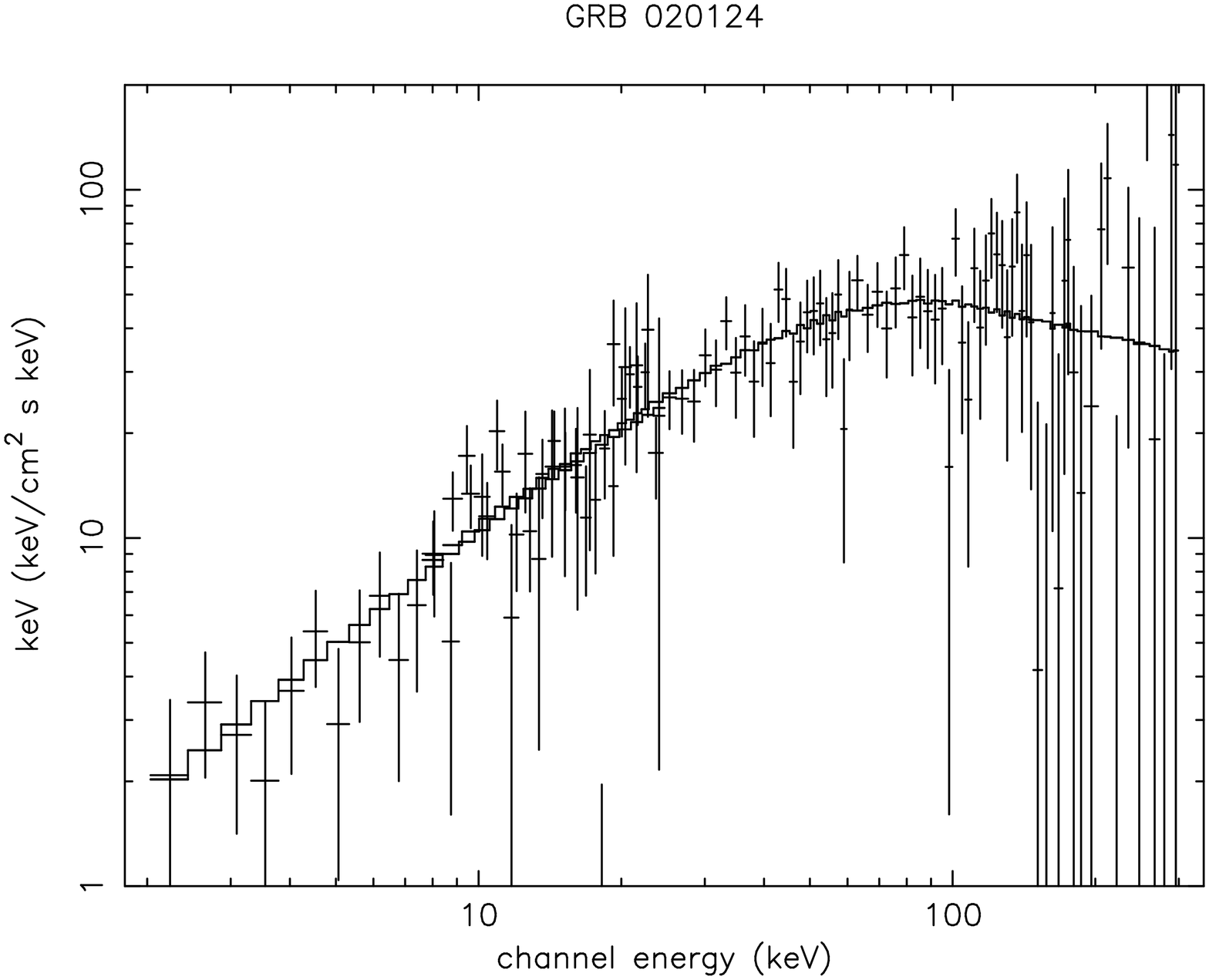}
\includegraphics[angle=270,scale=0.25]{f3b.eps}
\caption{Left panel: Unfolded energy spectrum of GRB 020124 in the range 2-400 keV.
Right panel: Unfolded energy spectrum of GRB 030323 in the range 2-400 keV. 
\label{figsp}}
\end{figure}

\begin{figure}
\plotone{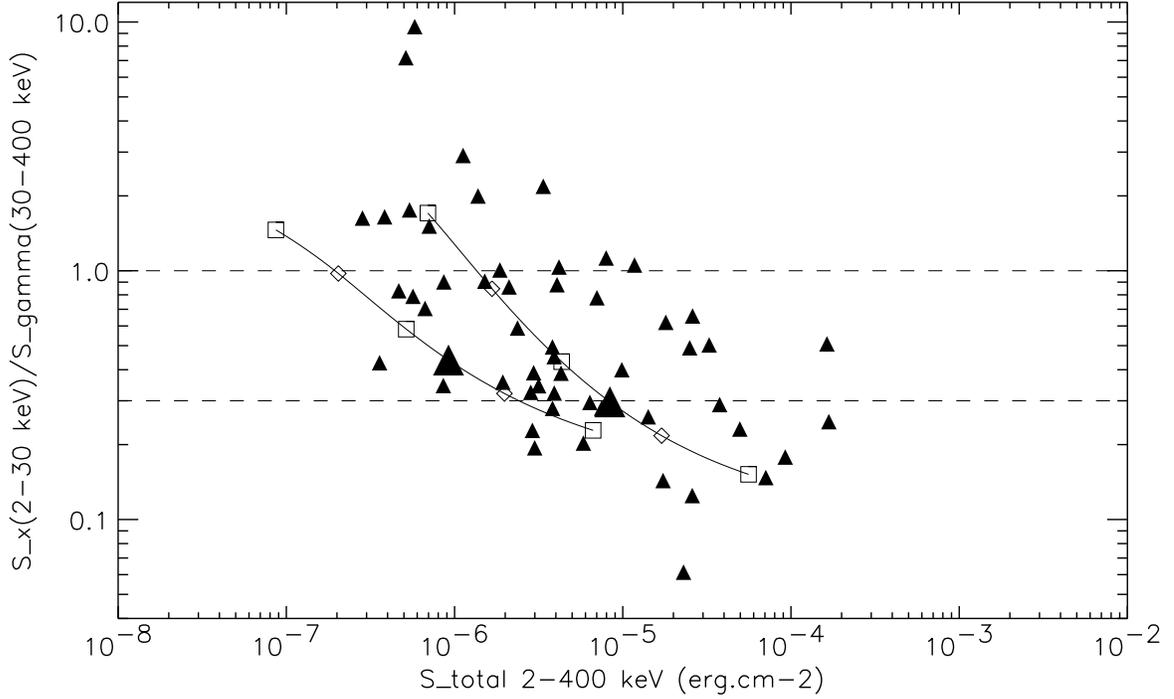}
\caption{The position of GRB 020124 (rightmost large triangle) and GRB 030323
(leftmost large triangle) in a fluence-softness plot, among other GRBs detected
with HETE (Barraud et al. 2004). The two solid lines show the tracks of 
GRB 020124 and GRB 030323 with 
redshift, from z=1 to z=20. Open squares indicate redshifts 1, 5 and 20 and 
open diamonds redshifts 2 and 10. This figure shows that at a redshift of unity 
GRB 020124 would have been in the bright end of the GRB fluence distribution,
and GRB 030323 would have been in the middle of the distribution.
This figure also shows that GRB 030323 is at the lower boundary of the fluence
distribution of HETE GRBs, while GRB 020124 would still lie in the middle of this
distribution, even at redshift z$\sim$10. An estimate of the maximum
redshift at which \grba\ could have been detected by HETE is given in section 
\ref{dmax}. The upper dashed horizontal line shows the boundary between
X-Ray Flashes (above the line) and X-ray rich GRBs. The lower dashed 
horizontal line shows the boundary between
X-ray rich GRBs and normal GRBs (below the line).  
\label{softness}}
\end{figure}

\end{document}